**Assessment of deep learning models integrated with weather and environmental variables for wildfire spread prediction and a case study of the 2023 Maui fires**

Jiyeon Kim [a], Yingjie Hu [a, b], Negar Elhami-Khorasani [c], Kai Sun [a] , and Ryan Zhenqi Zhou [a]

[a.] *GeoAI Lab, Department of Geography, University at Buffalo, Buffalo, NY, USA*

[b.] *Department of Computer Science and Engineering, University at Buffalo, Buffalo, NY, USA*

[c.] *Department of Civil, Structural and Environmental Engineering, University at Buffalo, NY, USA*

**Abstract:** Predicting the spread of wildfires is essential for effective fire management and risk assessment. With the fast advancements of artificial intelligence (AI), various deep learning models have been developed and utilized for wildfire spread prediction. However, there is limited understanding of the advantages and limitations of these models, and it is also unclear how deep learning-based fire spread models can be compared with existing non-AI fire models. In this work, we assess the ability of five typical deep learning models integrated with weather and environmental variables for wildfire spread prediction based on over ten years of wildfire data in the state of Hawaii. We further use the 2023 Maui fires as a case study to compare the best deep learning models with a widely-used fire spread model, FARSITE. The results show that two deep learning models, i.e., *ConvLSTM* and *ConvLSTM with attention*, perform the best among the five tested AI models. FARSITE shows higher precision, lower recall, and higher F1-score than the best AI models, while the AI models offer higher flexibility for the input data. By integrating AI models with an explainable AI method, we further identify important weather and environmental factors associated with the 2023 Maui wildfires.

## 1 Introduction

Wildfire disasters have increased in frequency and intensity in recent years. They devastated communities and ecosystems across the globe, from Australia and the Arctic to North and South America (Miezïte et al., 2022; Sullivan, A. et al., 2022; Noi et al., 2025). Research suggests that climate change will likely exacerbate drought conditions in susceptible regions in the coming years, which can result in an increase of dry and dead plant matter and can prolong and intensify wildfires (Fettig et al., 2013; Nones et al., 2024; Stephens et al., 2018). Predicting the spread of wildfires is an important approach for managing wildfires and mitigating their negative impacts.

A variety of wildfire spread models have been developed, which can be organized in a spectrum ranging from purely physical models to purely empirical models (Sullivan, 2009a, 2009b, 2009c). Purely physical models are those that completely rely on fundamental principles of physics to predict fire spread. For example, the Wildland-Urban Interface Fire Dynamics

Simulator (WFDS) (Mell et al., 2007) is a physics-based model that explicitly solves the governing equations of combustion, heat transfer, and fluid dynamics to predict the behavior of wildfires. While built on good understandings of physics, purely physical models are often computationally prohibitive and can only run in small-scale laboratory settings or have to be largely simplified for real-world applications (Sullivan, 2009a). Purely empirical models are those that are developed completely based on empirical fire spread data from field observations or laboratory experiments. For example, Fernandes (2001) developed a purely empirical fire spread model based on fire data and weather data collected from field experiments and prescribed burns conducted in different types of shrub in Portugal. Purely empirical models have good pragmatic values as they can often run rapidly in real-world settings (Sullivan, 2009b); however, their fire spread predictions can be difficult to interpret due to the lack of clear physical principles, and their applicability can be limited to cases similar to those whose empirical data were used for developing the models. Between purely physical and purely empirical models, there are hybrid fire spread models that make use of both physical principles and empirical data. Examples include FARSITE (Finney, 1998), FSim (Finney et al., 2011), and WRF-Fire (Coen et al., 2013), which combine physical representations of fire spread, such as Rothermel's equation and Huygens' wave principle, with empirical and observed fire and weather data.

With the fast advancements of artificial intelligence (AI), many deep learning models have been developed for wildfire science and management (Jain et al., 2020). Deep learning models can be considered as a type of purely empirical models and have inherited the corresponding characteristics. Deep learning fire spread models are straightforward to implement and train on empirical fire spread data. While the training process can take some time, a trained model can usually make predictions rapidly. With the ability to model complex and nonlinear relationships among data variables, deep learning models have also shown promising performance in fire spread predictions in recent studies (Liang et al., 2019; Jiang et al., 2023; Burge et al., 2023; Bhowmik et al., 2023). The predictions of deep learning models can be hard to explain; meanwhile, recent development of explainable AI methods offers new possibilities to increase model explainability (Zakari et al., 2025).

A number of studies have leveraged deep learning to predict wildfire spread. Liang et al., (2019) trained a long short-term memory (LSTM) model using time series data to capture temporal trends in fire occurrences. Khennou & Akhloufi (2023) used U-Net (Ronneberger et al., 2015) to extract spatial patterns of wildfire spread by framing the problem as an image segmentation task where each cell is classified as either 'fire' or 'no-fire'. Burge et al., (2023) leveraged the Convolutional LSTM (ConvLSTM) model (Shi et al., 2015) to predict the dynamics of wildfire spread over successive time steps. Some studies also added the attention mechanism to U-Net and ConvLSTM, which allows the models to concentrate on relevant spatial regions or time steps of the data (Fitzgerald et al., 2023; Masrur et al., 2024). While different deep learning models have been developed and utilized, there is limited understanding of their advantages and limitations in fire spread prediction. It is also unclear how deep learning-based fire spread models can be compared with existing non-AI models.

This study fills these two research gaps. Specifically, we compare five deep learning models typically used in the literature for fire spread prediction based on over a decade of wildfire data in the state of Hawaii. We also compare deep learning models with a widely used non-AI model FARSITE (Finney, 1998), which is a physics-informed and semi-empirical fire spread simulation model, in a case study of the 2023 Maui fires. Among the Maui fires in August 2023, the Lahaina fire claimed more than 100 lives and was considered as the deadliest wildfire in the United States for more than a century (O'Kane, 2023; Treisman, 2023). We ask three research questions (RQs) in this study:

RQ1: *What are the advantages and limitations of five different deep learning models typically used for fire spread prediction?*

RQ2: *What are the advantages and limitations of the deep learning models compared with a physics-informed and semi-empirical fire spread model?*

RQ3: *What are the major weather and environmental factors, such as wind, vegetation, and topography, associated with the 2023 Maui fires?*

The remainder of this paper is organized as follows. Section 2 describes the study area and materials. Section 3 presents our experimental design and describes the deep learning models and the FARSITE model used for experiments. Section 4 presents the obtained results, and Section 5 discusses these results and answers the RQs. Finally, Section 6 concludes this work.

## 2 Study area and materials

### *2.1 Study area and fire data*

Our study area is the state of Hawaii, and we will also conduct a case study specifically focusing on Maui for the 2023 Maui fires. Fire data used in this study were collected from the Fire Information for Resource Management System (FIRMS) of the National Aeronautics and Space Administration (NASA). The FIRMS data provides global fire occurrence information, in the form of point-based fire locations, detected by NASA Visible Infrared Imaging Radiometer Suite (VIIRS) on the Suomi National Polar-orbiting Partnership (NPP) satellite. The VIIRS images have a spatial resolution of 375 meters and a temporal resolution of about 12 hours. The FIRMS fire data were recorded as fire location points at the center of each grid cell of the VIIRS images where a fire is detected. We use the FIRMS fire data in the state of Hawaii from January 20, 2012 (which is the earliest date when the VIIRS data is available) to August 12, 2023, which covers a total time period of 11 years and 7 months. We only use fire points whose confidence is labeled as "*nominal*" or "*high*" in the FIRMS dataset to focus on high-quality data. Figure 1 shows the study area and the fire locations.

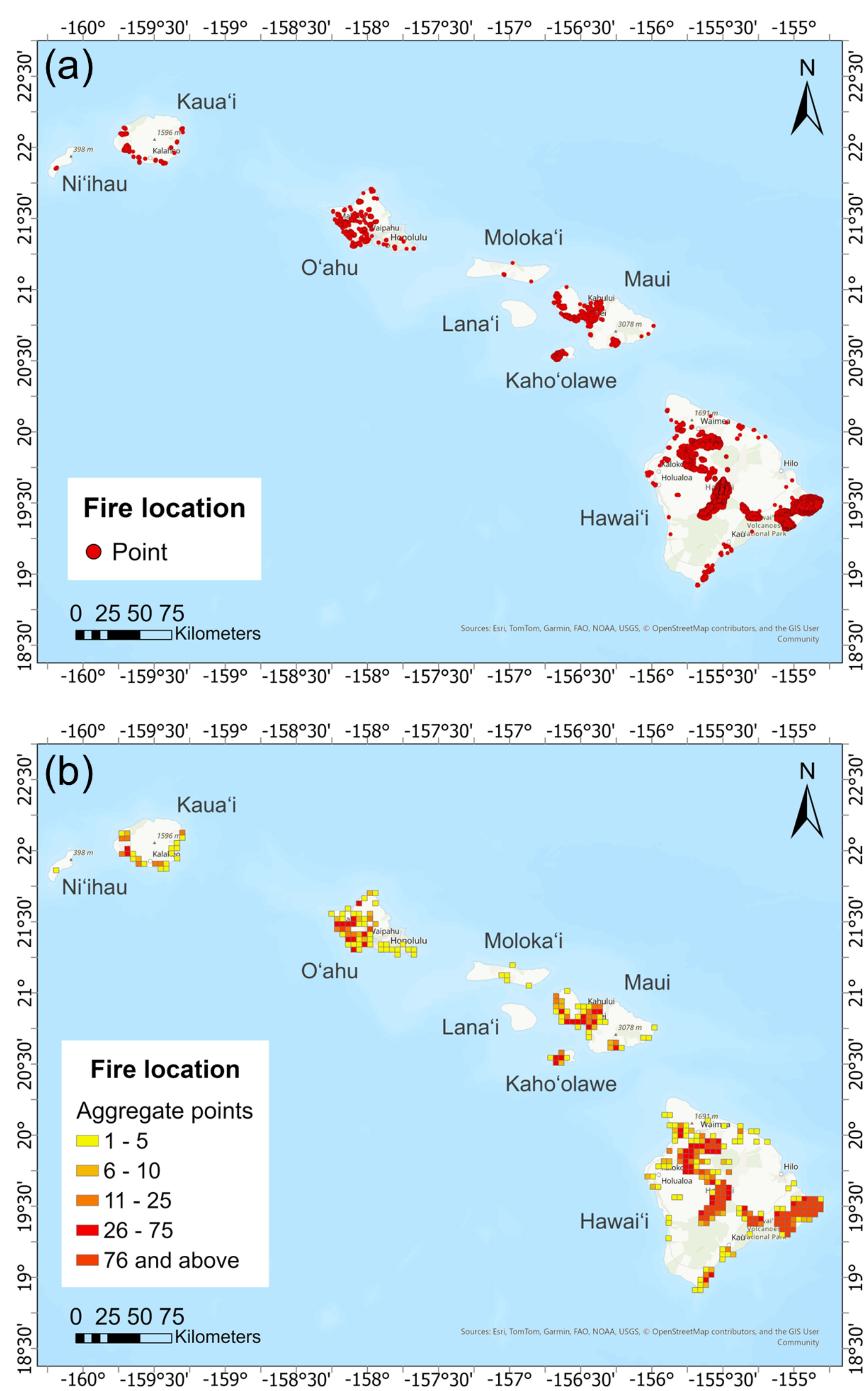

Figure 1. The study area and the NASA FIRMS fire location data: (a) raw fire location data; (b) fire location data aggregated to grid cells.

### *2.2 Environmental and weather data*

In addition to fire data, we use environmental and weather data commonly used in the literature as input features for deep learning models to predict fire spread (Gerard et al., 2023; Huot et al., 2022; Kondylatos et al., 2022; Marjani et al., 2023; Jiang et al., 2023). These data variables can be categorized into four groups: *weather*, *topography*, *vegetation,* and *anthropogenic activity*. Each variable represents an aspect of the environment and weather that may affect fire spread behaviors. Table 1 summarizes these variables and their data sources.

Table 1. Four categories of environmental and weather variables.

| **Category** | **Variable** | **Data Source** |
|---|---|---|
| Weather | Temperature | ERA-5 and ERA-5 Land |
| | Total precipitation | |
| | Relative humidity | |
| | Wind speed | |
| | Wind direction | |
| Topography | Elevation | United States Geological Survey |
| | Slope | |
| | Aspect | |
| Vegetation | Normalized difference vegetation index | NASA VIIRS Vegetation Indices (via Google Earth Engine) |
| Anthropogenic activity | Population density | LandScan Global |

- *Weather*: To capture changes in weather, we use data from ERA-5 and ERA-5 Land which provide hourly weather updates. Data for six variables, i.e., *temperature*, *total precipitation*, *relative humidity*, *wind speed*, and *wind direction*, are obtained. We primarily use ERA-5 Land data which has a spatial resolution of about 9 kilometers; in the case of missing data, ERA-5 data is used to fill the data gap, which has a spatial resolution of about 31 kilometers. The models are provided with weather data prior to the prediction time steps.
- *Topography*: For topographic variables, we use the National Elevation Dataset for the state of Hawaii from the United States Geological Survey (USGS). The data has a spatial resolution of 10 meters, and we derive three variables from the data: *elevation*, *slope*, and *aspect*.
- *Vegetation*: For vegetation, we use the Normalized Difference Vegetation Index (NDVI) data. Specifically, we use the NDVI from the NASA VIIRS Vegetation Indices (VNP13A1) dataset. The original data provides NDVI at a spatial resolution of 500 meters and a temporal resolution of eight days. We obtain the data for our study time period using Google Earth Engine (GEE). Specifically, we obtain the daily composite NDVI data from GEE using a 16-day window to obtain pixel-wise median of available NDVI observations in the past 16 days. NDVI has been frequently used in previous deep learning-based fire spread prediction research for capturing vegetation characteristics (Huot et al., 2022; Fitzgerald et al., 2023; Shadrin et al., 2024). When comparing deep learning models with the FARSITE model later, we will use vegetation and fuel variables that are typically used in traditional fire spread models, such as fuel model, canopy height, and canopy cover.

- *Anthropogenic activity:* We use population density to partially capture anthropogenic activity. The population density data used in this study is obtained from the LandScan Global Population Database developed by the Oak Ridge National Laboratory (ORNL). The data has a spatial resolution of 1 kilometer and is updated annually. We obtain yearly LandScan data from 2012 to 2023.

## 3 Method

### *3.1 Overview of research design*

We will compare the performances of five different deep learning models and the FARSITE model on the task of fire spread prediction. We formalize this task as below: *given the initial fire locations and the current weather and environmental conditions, predict the fire locations at future time steps.* Specifically, we will compare model predictions in four future time steps, i.e., 12 hours later, 24 hours later, 36 hours later, and 48 hours later. An overview of our research design is provided in Figure 2. It consists of three main steps: (1) data preprocessing, (2) model training, and (3) model comparison. For step (1), we preprocess the NASA FIRMS fire location data to extract individual fires based on the spatial and temporal continuity of the fire location data. For step (2), we train five different deep learning models for fire spread prediction using historical fire data, environmental data, and weather data. For step (3), we first compare the performances of the deep learning models based on the Hawaii fire data (excluding the 2023 Maui fires), and then compare the best deep learning models with the FARSITE model on the 2023 Maui fire data. In the following, we discuss each of the three steps.

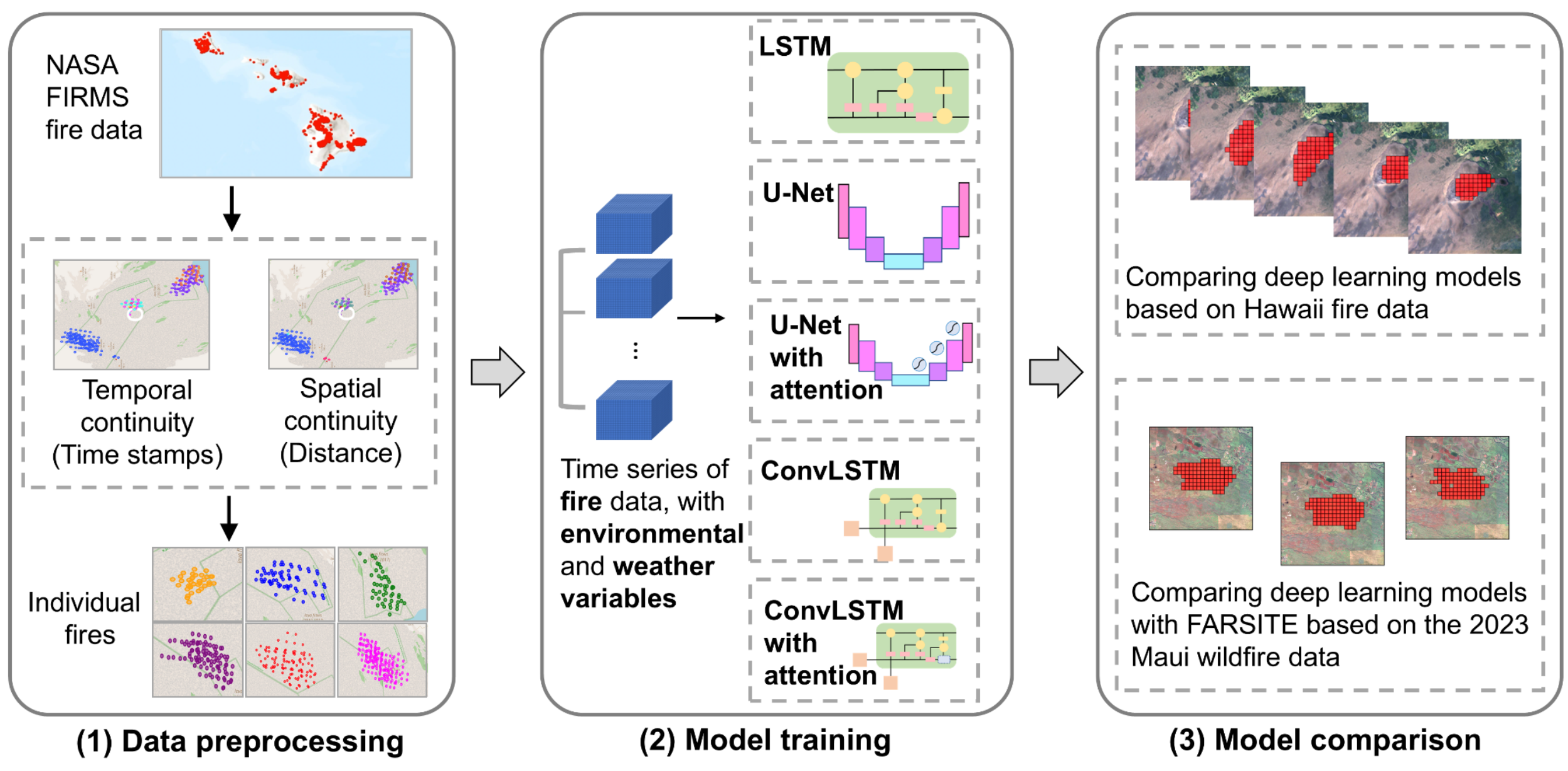

Figure 2. An overview of our research design.

*3.2 Data preprocessing*

The raw NASA FIRMS data have location points of different fires all combined together. Since our goal is to predict the spread of individual fires, our first step is to preprocess the raw FIRMS data to extract individual fires. To do so, we separate the fire locations of individual fires based on their spatial and temporal continuity. For spatial continuity, we leverage the Density-Based Spatial Clustering of Applications with Noise (DBSCAN) clustering algorithm by setting the minimum number of points required for a cluster to 3 and the search radius to 530.25 meters (which is 375 meters * 1.414). We use 530.25 meters as the search radius because the horizontal and vertical distances between two fire location points are about 375 meters; thus, using a search radius of 530.25 meters for DBSCAN allows the algorithm to reach all nearby points in horizontal, vertical, and diagonal directions in the spatial clustering process. For temporal continuity, we utilize the time stamps of fire locations to ensure that the points of the same fire are temporally continuous with a time interval of about 12 hours (since the VIIRS images have a temporal resolution of about 12 hours). This spatial and temporal data preprocessing is done using the Density-based Clustering tool in the ArcGIS Pro (version 3.4.0) software. With the points of individual fires obtained, we further split the points into different time steps using an interval of about 12 hours apart, and put the fire points of the same time into the same time step. Points at the same time step are further converted into a raster image with a spatial resolution of 375 meters. Pixel values are set to 1 for fire locations and 0 for non-fire locations on the raster images. The raster images are prepared with a size of 64 * 64 pixels, which covers a 24 * 24 kilometer square with a total area of 576 $km^2$ (about 142,333 acres). This size can cover all fires in the data and leave a sufficient buffer distance for even the largest fires for spread prediction. Figure 3 illustrates these steps in preprocessing the fire location data.

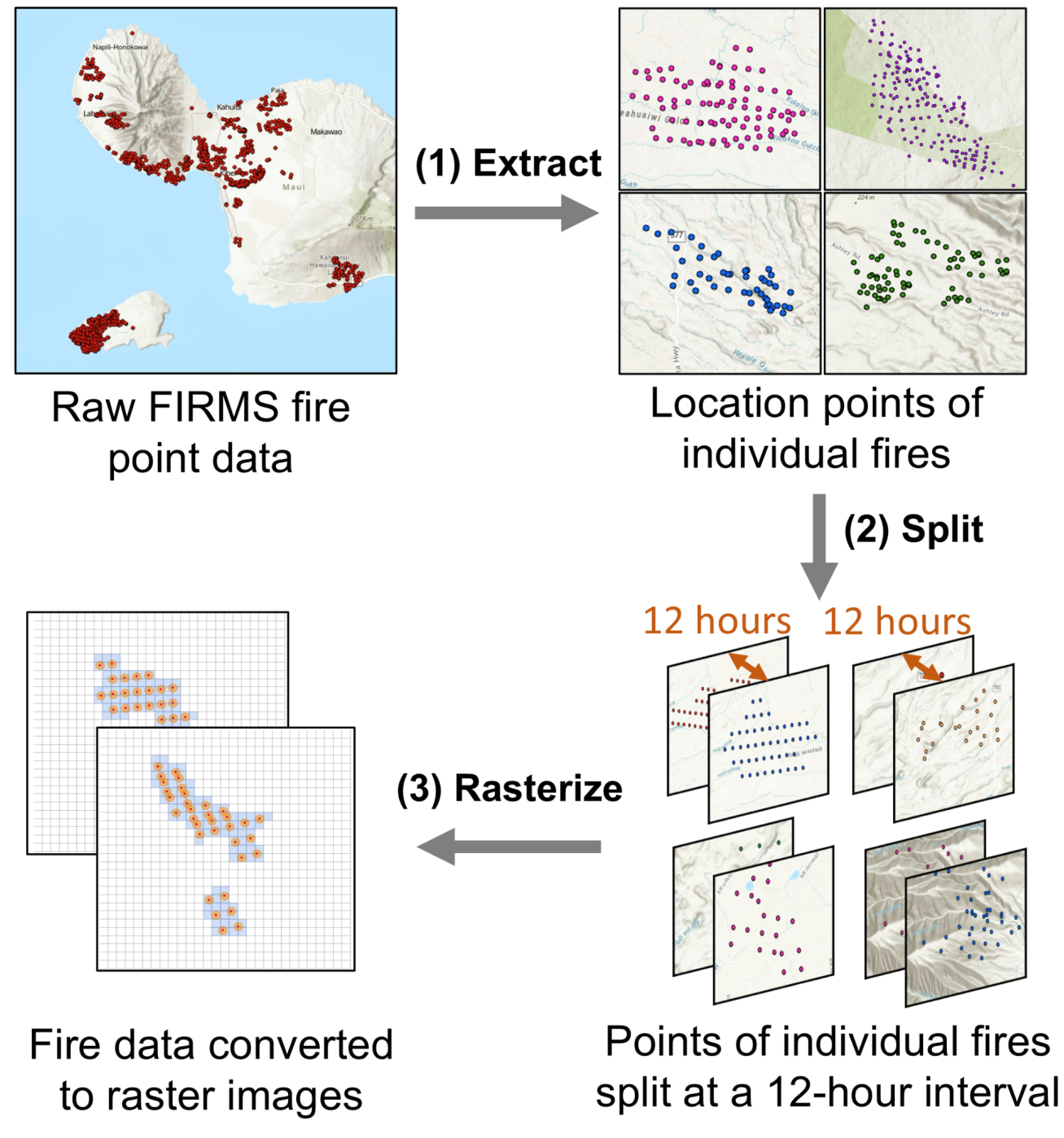

Figure 3. Major steps involved in preprocessing the fire location data.

With the extracted individual fires, we use those that have at least eight time steps (i.e., the fire lasted for at least four days) to train and test deep learning models. In total, this dataset has 221 distinct fire events covering the entire state of Hawaii from January 2012 to August 2023. The four 2023 Maui fires are not included in this dataset, which are used for the case study later. The minimum burnt area of the 221 wildfires is 1.55 km² (about 383 acres); the maximum burnt area is 221.48 km² (about 54,729 acres); and the average burnt area is 9.80 km² (about 2,421 acres). A summary of the monthly distribution of these wildfires is provided in Supplementary Figure S1.

In addition to the fire data, all environmental and weather variables are resampled to align with the spatial and temporal resolutions of fire data. For topographic data whose spatial resolution is finer than 375 meters, we aggregate data values to a 375-meter grid using the mean values. For weather, vegetation, and anthropogenic activity data whose spatial resolution is coarser than 375 meters, we use cubic function to downscale the data to 375 meters. In terms of temporal resolution, the weather data is updated hourly and we aggregate them to 12 hours by taking both the minimum and maximum values of each weather variable following the literature (Huot et al., 2022; Kondylatos et al., 2022). For the other variables, we use the temporally closest data based on the time of the fire. For example, the population density data in 2022 are used for fires in the year of 2022, since the data are updated annually.

*3.3 Model training*

With the input data prepared, we train five deep learning models used in the literature for fire spread prediction. Brief descriptions of these models are provided below:

- LSTM: The LSTM model is a variant of the Recurrent Neural Network (RNN) model, and is particularly strong in handling time-series data. The LSTM model is designed to address issues related to long-term temporal dependency. The main component of the LSTM model is the LSTM cell which contains three gates: input gate, output gate, and forget gate. Together, these three gates regulate the flow of the information to the model at each time step, and help the model retain important information over extended time periods. The architecture of the LSTM model used in this study follows the model developed by Kondylatos et al. (2022). Since the input data of LSTM needs to be in 1D format, we flatten the 2D images of individual fires before feeding them to the model.

- U-Net: U-Net is a special type of CNN which consists of three main components: the encoder block, the decoder block, and the skip connection. The encoder block reduces the dimensions of the input image by capturing important image features. The decoder block progressively reconstructs an image as the output result. Skip connections directly transfer high-resolution image information from the encoder to the decoder, and enable the model to preserve and utilize some spatial details of the image that might be lost during the decoding process. The architecture of the U-Net model follows Fitzgerald et al. (2023). Since U-Net can directly take a 2D image as its input, we do not need to flatten the fire images as needed for LSTM.

- U-Net with attention: This model uses the same architecture of U-Net but adds the attention mechanism (Fitzgerald et al., 2023), which enhances the model's ability to focus on some critical parts of the input image. The attention mechanism involves three parts: Query, Key, and Value operations. Query represents the current analysis of the model, while Key represents different parts of the input data. Similarities between Query and Key are calculated to obtain weights that are then combined with Value to determine the final attention scores. The attention mechanism enables U-Net to selectively focus on important regions in the image, and potentially improves the ability of the model for fire spread prediction.

- ConvLSTM: ConvLSTM is a neural network architecture that combines the features of LSTM and CNN models, and is particularly useful in handling time-series data which have images at each time step. While LSTM captures temporal dependency of the data, it does not directly model the spatial structure of the input data (since the input data is in 1D format). ConvLSTM adds convolutional operations to the LSTM model, and allows the model to directly process a 2D image at each time step. As a result, ConvLSTM captures both spatial and temporal dependencies of the data, which could help increase the accuracy of fire spread prediction. The architecture of the ConvLSTM model used in this study follows Masrur & Yu (2023).

- ConvLSTM with attention: This model uses the same architecture of ConvLSTM but adds the attention mechanism (Masrur & Yu, 2023). Similar to U-Net, the attention mechanism calculates the similarity between Query and Key, dynamically adjusts the weights for Value, and helps the model to focus on the most relevant temporal and spatial features in the data. As a result, ConvLSTM with attention may be able to provide more accurate predictions than the original ConvLSTM model in spatiotemporal tasks such as fire spread prediction.

We train these deep learning models using wildfire data in the state of Hawaii from 2012 to 2023, excluding the data of the 2023 Maui fires. We use 70% of the Hawaii wildfire data for training, 10% for validation, and 20% for testing. All fire data are organized into four input time steps and four output time steps, and this data organization enables the trained models to predict four time steps into the future, i.e., fire spread in 12 hours, 24 hours, 36 hours, and 48 hours. The LSTM model and the two ConvLSTM models are directly trained on this organized fire dataset. The two U-Net models can only predict one step ahead at a time, and we use an iterative training approach where the first time step prediction is based on the initial input and each subsequent prediction is based on the previously predicted output. While the training process uses four input time steps, all trained models can take fewer input steps to make predictions. For example, a trained model can take only one step of the current fire locations and make predictions for the next four steps. The training process utilizes early stopping to prevent overfitting, and the training process stops when the performance of the model ceases to improve on the validation data.

*3.4 Model comparison*

We conduct model comparison in two groups of experiments. First, we compare the performance of the five deep learning models. We use the test dataset from the Hawaii wildfire data, i.e., 20% of the Hawaii wildfire data from January 2012 to August 2023, to compare the five deep learning models. This allows us to evaluate the performance of these models more comprehensively based on the many fires in the test dataset. Second, we compare the best deep learning models with the FARSITE model using data from the 2023 Maui fires. FARSITE makes predictions based on a variety of input datasets describing landscape and environmental conditions, such as fuel model, fuel moisture, canopy height, cloud cover, temperature, humidity, elevation, and wind direction. The 2023 Maui fire data have four individual fires: Olinda fire, Kula fire, Pulehu/Kihel fire, and Lahaina fire. We compare FARSITE with the best deep learning models for these four fires. From a practical perspective, FARSITE requires the model to be run manually for each individual fire, which is difficult to complete for many fires but can be done for the four individual fires in the 2023 Maui fires.

We use three evaluation metrics, i.e., *precision*, *recall*, and *F1-score*, to compare the performance of different models. The calculations of these three metrics are shown in Equations (1)-(3) :

$$Precision = \frac{|Correctly\ predicted\ fire\ pixels|}{|All\ predicted\ fire\ pixels|} \quad , \quad (1)$$

$$Recall = \frac{|Correctly\ predicted\ fire\ pixels|}{|All\ correct\ fire\ pixels|} \quad , \quad (2)$$

$$F1-score = 2 \times \frac{Precision \times Recall}{Precision + Recall} \quad . \quad (3)$$

To calculate the three metrics, we compare the predicted fire locations with the true fire locations in the next time steps. We conduct the comparison in both individual time steps and all time steps in a collective manner. A model with high precision but low recall suggests the fire locations predicted by the model are likely to be correct, although the model can miss some true fire locations (i.e., false negatives). A model with high recall but low precision suggests that the model is unlikely to miss true fire locations, but it might predict fire locations that are not true (i.e., false positive). A model with a high F1-score has a balanced performance in both precision and recall.

## 4 Results

### *4.1 Comparison of the five different deep learning models on the Hawaii fire data*

We first compare the performance of the five deep learning models based on the Hawaii test dataset. Each trained model is used to make predictions for four time steps, i.e., 12 hours later, 24 hours later, 36 hours later, and 48 hours later. Their overall performances from all four time steps are summarized in Table 2. As can be seen, the *ConvLSTM* model achieves the highest F1-score, which is slightly higher than the *ConvLSTM with attention* model. The *LSTM* model achieves the third highest F1-score, while the two *U-Net* models achieve the lowest F1-score largely due to their low precisions. *U-Net with attention* achieves the highest recall, and ConvLSTM achieves the highest precision. It seems that the attention mechanism generally increases recall, as shown in both the *U-Net* and *ConvLSTM* models.

Table 2. Overall performance of the five deep learning models across all four time steps.

| Model | Precision | Recall | F1-score |
|---|---|---|---|
| ***LSTM*** | 0.5307 | 0.4621 | 0.4940 |
| ***U-Net*** | 0.3102 | 0.6916 | 0.4283 |
| ***U-Net with attention*** | 0.3071 | **0.7463** | 0.4351 |
| ***ConvLSTM*** | **0.5603** | 0.6324 | **0.5942** |
| ***ConvLSTM with attention*** | 0.5397 | 0.6531 | 0.5910 |

We further show the performances of the models in each individual time step in Figure 4. We calculate the precision, recall, and F1-score metrics for each individual fire, and show both the average performance and the performance interquartile range (25th-75th percentile) of the models. In terms of precision, the two *ConvLSTM* models and the *LSTM* model achieve higher average precision than the two *U-Net* models. In terms of recall, the two *U-Net* models achieve the highest average recall, and the *LSTM* model has the lowest recall among the tested models. In terms of F1-score, the two *ConvLSTM* models achieve the highest average F1-score in each of the four time steps, demonstrating a balanced overall performance.

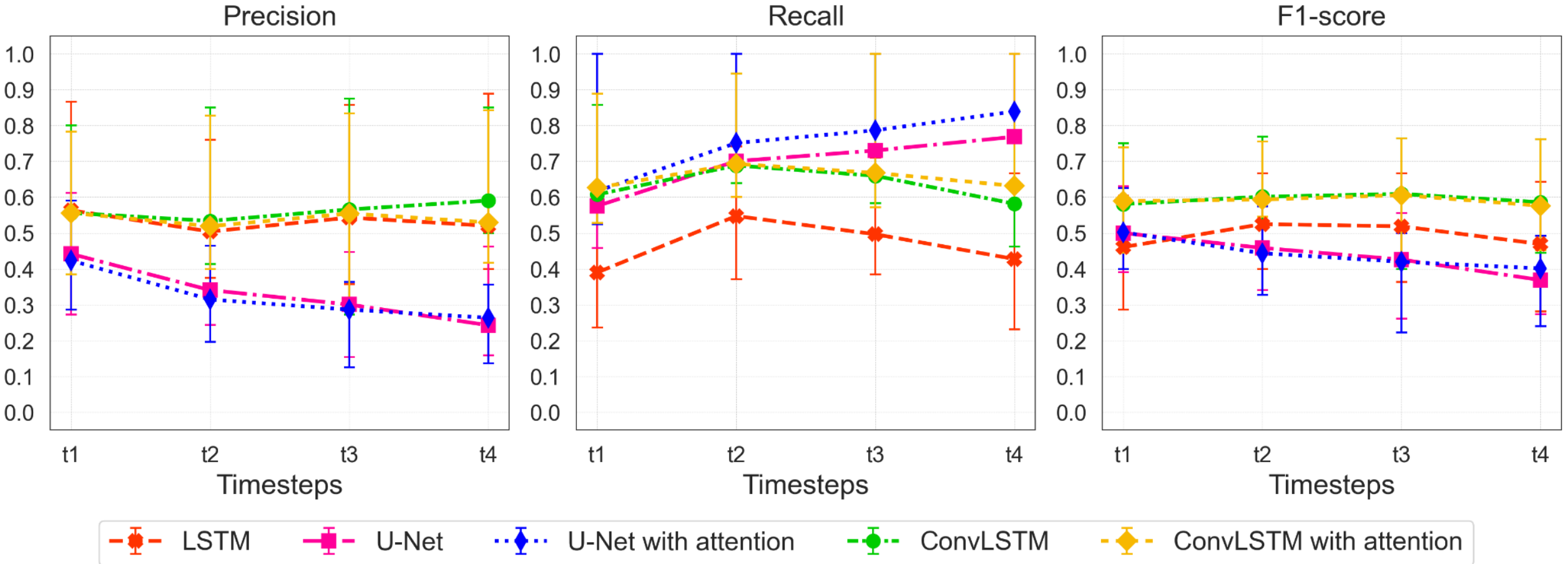

Figure 4. Performance of the five different deep learning models in different time steps.

Given that the average performance of some models is close and their interquartile ranges overlap, we further conduct pairwise Nemenyi post-hoc tests for each model pair for all time steps. The results are summarized in Figure 5, with darker colors indicating lower $p$ values and more significant differences. In terms of precision, the two *ConvLSTM* models and the *LSTM* model show significant differences compared with the two *U-Net* models. In terms of recall, the two *ConvLSTM* models and the two *U-Net* models show significant differences from the *LSTM* model. In terms of F1-score, the two ConvLSTM models show significant differences from the other models, although the difference between the two ConvLSTM models is not significant.

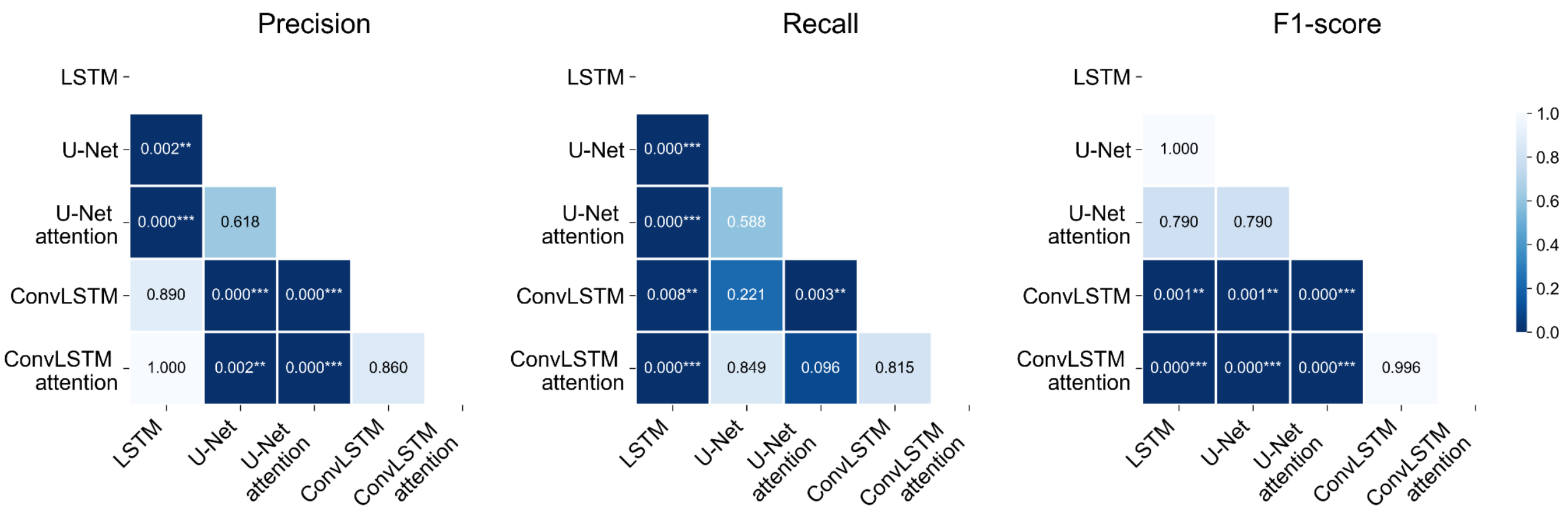

Figure 5. Pairwise Nemenyi test p-values for precision, recall, and F1-score (* p < 0.05; ** p < 0.01; *** p < 0.001)

We also visualize the predictions of the five deep learning models. Figure 6 shows one example, with the true fire locations shown in the first row and the predicted fire locations from different models shown in the following rows.

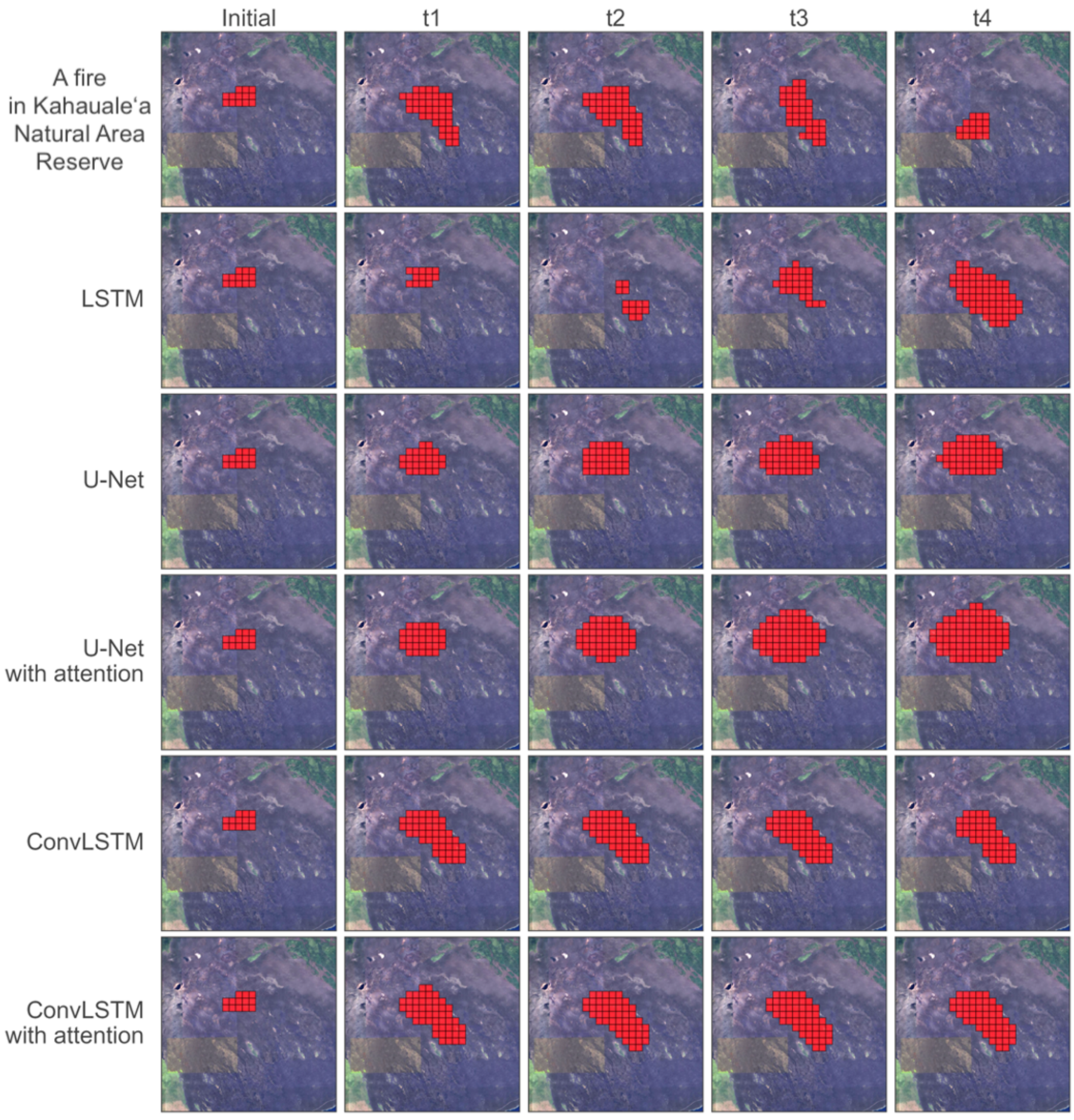

Figure 6. Visualization of the true fire locations and predicted locations from different models.

One interesting observation is that the true fire locations do not follow a purely expanding pattern, but can also shrink as fires in some locations burn out (e.g., the true fire locations at the time step $t_4$). While this dynamic change is captured to some extent by the *LSTM* model and the *ConvLSTM* models, the two *U-Net* models, especially the *U-Net with attention* model, seem to predict a mostly expanding pattern. This expanding pattern explains the higher recalls but lower precisions of the *U-Net* models, since such a pattern is unlikely to miss a true fire location and will likely include many false positives. The predicted fire locations of the *LSTM* model seem to change more across the four time steps than the predictions of the two *ConvLSTM* models, although more changes do not mean higher prediction accuracy. While we only include one visualization example in Figure 6 due to space limitation, similar prediction patterns are also observed in other fires. We include two more examples in Supplementary Figure S2.

To understand the computational costs of different models, we summarize the training and test time of the five models in Table 3. All time values are recorded on the same Dell 7920 server with two Intel Xeon Silver 4116@ 2.10 GHz CPU (12 cores) and two NVIDIA Quadro P5000 GPUs (16 GB VRAM each). Each model is trained and tested five times and the average time is reported in Table 3. For all five models, they are highly efficient in terms of making predictions after training, with all test times being under one second. Model training time varies, ranging 41.1 seconds by the *U-Net with attention* model to 1056.1 seconds (i.e., 17.6 minutes) by the *ConvLSTM with attention* model.

Table 3. Training and test time of the five deep learning models (in seconds).

| Model | Training time | Test time |
|---|---|---|
| ***LSTM*** | 158.6 | **0.066** |
| ***U-Net*** | 64.9 | 0.471 |
| ***U-Net with attention*** | **41.1** | 0.695 |
| ***ConvLSTM*** | 629.2 | 0.308 |
| ***ConvLSTM with attention*** | 1056.1 | 0.977 |

*4.2 Comparison with FARSITE on the 2023 Maui fire data*

We further compare the deep learning models with FARSITE. The comparison is conducted based on the 2023 Maui fire data consisting of four fires: the Olinda Fire, the Kula Fire, the Pulehu-Kihei fire, and the Lahaina Fire. We select two deep learning models, *ConvLSTM* and *ConvLSTM with attention*, which have demonstrated the best overall performances in previous experiments for this comparison. There are two additional aspects that require our special consideration in order to conduct the comparison experiments. First, as a specialized fire spread model, FARSITE has its own required input data that are different from the typical input data used by deep learning models in the literature. For example, FARSITE requires the use of fuel moisture data which are not used in typical deep learning models. In addition, while NDVI is frequently used as an important input variable by deep learning models, FARSITE uses more detailed vegetation data, such as fuel model, canopy height, and canopy base height. With this consideration, we conduct two groups of experiments using both the original input data that we used in previous experiments as well as new input data based on the specific data requirements of FARSITE. Second, models may show different prediction performances in shorter and longer time steps. To understand model performance more comprehensively, we compare the deep learning models and FARSITE in both a single-step prediction experiment setting and also a multi-step prediction setting.

We first prepare the input data required by FARSITE, as summarized in Table 4. We use the same weather data from ERA-5 and ERA-5 Land but also add cloud cover data as required by FARSITE. We obtain topographic data from the USGS and vegetation data from LANDFIRE. For fuel moisture data, we download fire weather observation data from weather stations, and then use the software FireFamilyPlus (Bradshaw & McCormick, 2000), which is based on the National Fire Danger Rating System (NFDRS), to create the data. With the input data prepared, we run comparison experiments on both the original data used by previous deep learning models (i.e., data in Table 1) and also the new input data prepared specifically for FARSITE.

Table 4. Input data prepared based on the requirements of FARSITE.

| **Category** | **Variable** | **Data Source** |
|---|---|---|
| Weather | Temperature | ERA-5 and ERA-5 Land |
| | Precipitation | |
| | Relative humidity | |
| | Wind speed | |
| | Wind direction | |
| | Cloud cover | |
| Topography | Elevation | United States Geological Survey |
| | Slope | |
| | Aspect | |
| Vegetation | Fuel model | LANDFIRE |
| | Canopy height | |
| | Canopy cover | |
| | Canopy base height | |
| | Canopy bulk density | |
| Fuel | 1 hour | National Wildfire Coordinating Group |

| Category | Variable | Data Source |
|---|---|---|
| moisture | 10 hours | (NWCG) |
| | 100 hours | |
| | 1000 hours | |
| | Woody | |
| | Herbaceous | |

Figure 7 shows the comparison results of the two deep learning models and the FARSITE model on the 2023 Maui fire data. In Figure 7, each of the three rows shows the performance of the models in terms of precision, recall, and F1-score, respectively. The first column shows experiment results based on the original data input. Because the original data are insufficient for running the FARSITE model, the first column contains results of only the *ConvLSTM* and *ConvLSTM with attention* models. The second column shows results based on the input data prepared for FARSITE, and the two deep learning models are re-trained and tested using the same input data as used for FARSITE.

We can understand the effects of different input data on model performance by comparing the results in the two columns of Figure 7. Overall, using the input data required by FARSITE (right column) achieves a slightly higher precision, slightly lower recall, and slightly higher F1-score than using the original input data (left column). The FARSITE-based input data contain more details about vegetation, such as canopy height and fuel model, and fuel moisture, which likely have contributed to this performance difference. From a different perspective, this result also shows that using the original and simpler input data (e.g., using NDVI to represent vegetation) achieves only a slightly lower prediction accuracy for the deep learning models.

We can understand the performance difference of single-step and multi-step predictions by comparing the two groups of bars in each subplot in Figure 7. For single-step prediction, a model is asked to make a prediction for one step (12 hours later) based on one initial input step. If a fire has multiple steps, then the model is given the fire locations of each step and asked to predict the next step (e.g., given $t_1$ and predict $t_2$, and given $t_2$ and predict $t_3$). For multi-step prediction, a model is only given the first step of the fire, and it is asked to predict all the remaining steps in one shot. Overall, the results in Figure 7 show that when the models move from single-step prediction to multi-step prediction, they generally show decreased recall and F1-score although they mostly maintain precision. This result is understandable given that multi-step prediction is a more difficult problem than single-step prediction.

We can also understand the performance difference of the two deep learning models and the FARSITE model by focusing on the second column of Figure 7. In both single-step and multi-step predictions, FARSITE shows a higher precision, lower recall, and higher F1-score than the two other deep learning models. This result suggests an overall better performance of FARSITE (based on its higher F1-score), although FARSITE also seems to miss some true fire locations as reflected in its lower recall.

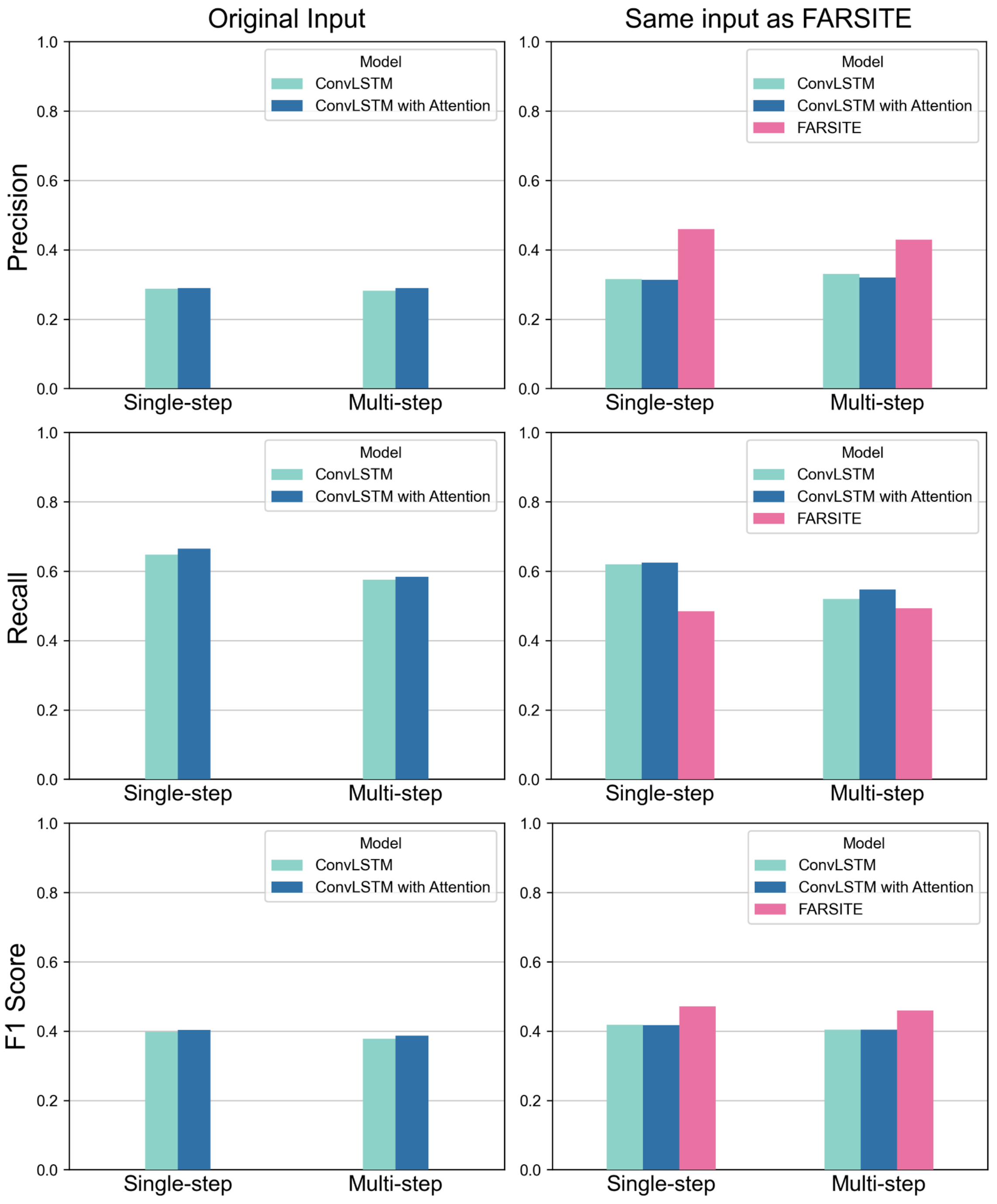

Figure 7. Comparison of the best deep learning models and the FARSITE model for predicting fire spread on the 2023 Maui fire data.

We further visualize the results of single-step prediction in Figure 8 and multi-step prediction in Figure 9. Both predictions are based on FARSITE-specific input data, so that we can show the results of all models. While the predicted fire locations are similar to the true locations in some steps, there are also large differences in some other steps. In particular, the Lahaina fire (last row in Figure 8 and subfigure (b) in Figure 9) has largely reduced burning locations within 12 hours (one time step), and the burning area of Pulehu/Kihel fire is also largely reduced in time step $t_2$ (second last row in Figure 8 and (d) in Figure 9). Model predictions of these two fires largely maintain the initial fire locations, leading to a large discrepancy between the model predictions and true fire locations. The Lahaina fire spread into urban areas, encountered some firefighting activities, and burnt houses and urban infrastructures; the Pulehu/Kihel fire was being actively fought during that time. Data about firefighting activities and detailed urban features, however, are not available to the models, which may have resulted in this large discrepancy between the predicted and true fire locations. One special case that should be noted is the prediction of FARSITE for the Lahaina fire. While it seems that FARSITE has successfully predicted the reduced fire locations at first glance, this result is, in fact, due to the inability of FARSITE to run fire spread simulation on the urban lands of Lahaina (Juliano et al., 2023). Thus, this limitation of FARSITE accidentally produces a better prediction in this special case of the Lahaina fire. If we remove the special case of Lahaina from evaluation, the two deep learning models and FARSITE archive very similar performance, with F1-scores of 0.4848, 0.4841, and 0.4883 for *ConvLSTM*, *ConvLSTM with attention*, and FARSITE, respectively for single-step prediction, and F1-scores of 0.4798, 0.4718, and 0.4743, respectively for multi-step prediction. Nevertheless, these performance scores are based on four fires only, and experiments on additional fires can help understand model performance more comprehensively.

In addition to prediction accuracy, we also compared the computational cost of the deep learning models and FARSITE. The deep learning models require some training time but can make predictions within a second. While FARSITE does not require training time, it takes about one minute to output predictions. The training and test times are summarized in Supplementary Table S1. Overall, these times seem to be negligible compared with the data preparation time.

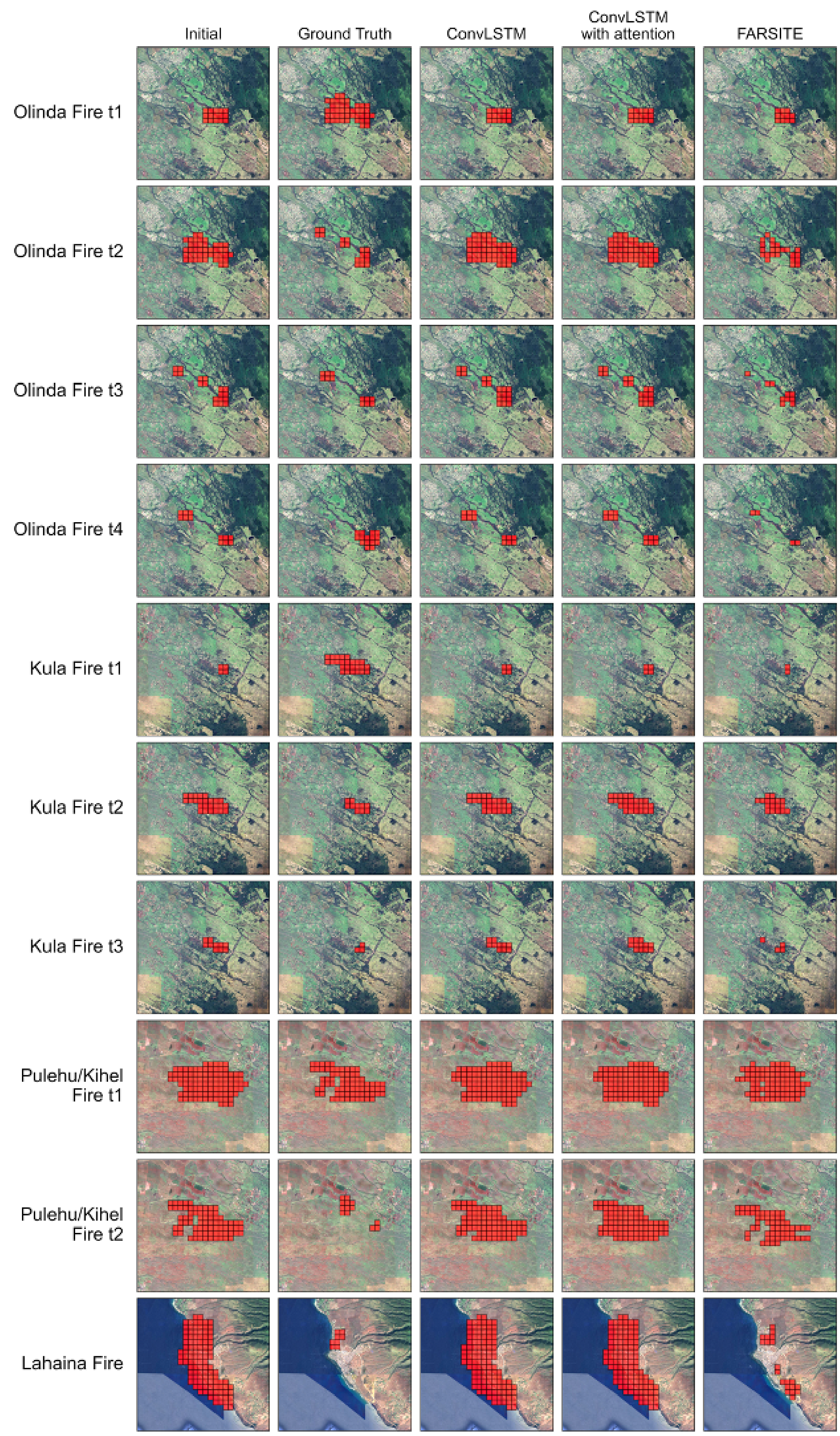

Figure 8. Single-step prediction results of all models based on the 2023 Maui wildfires.

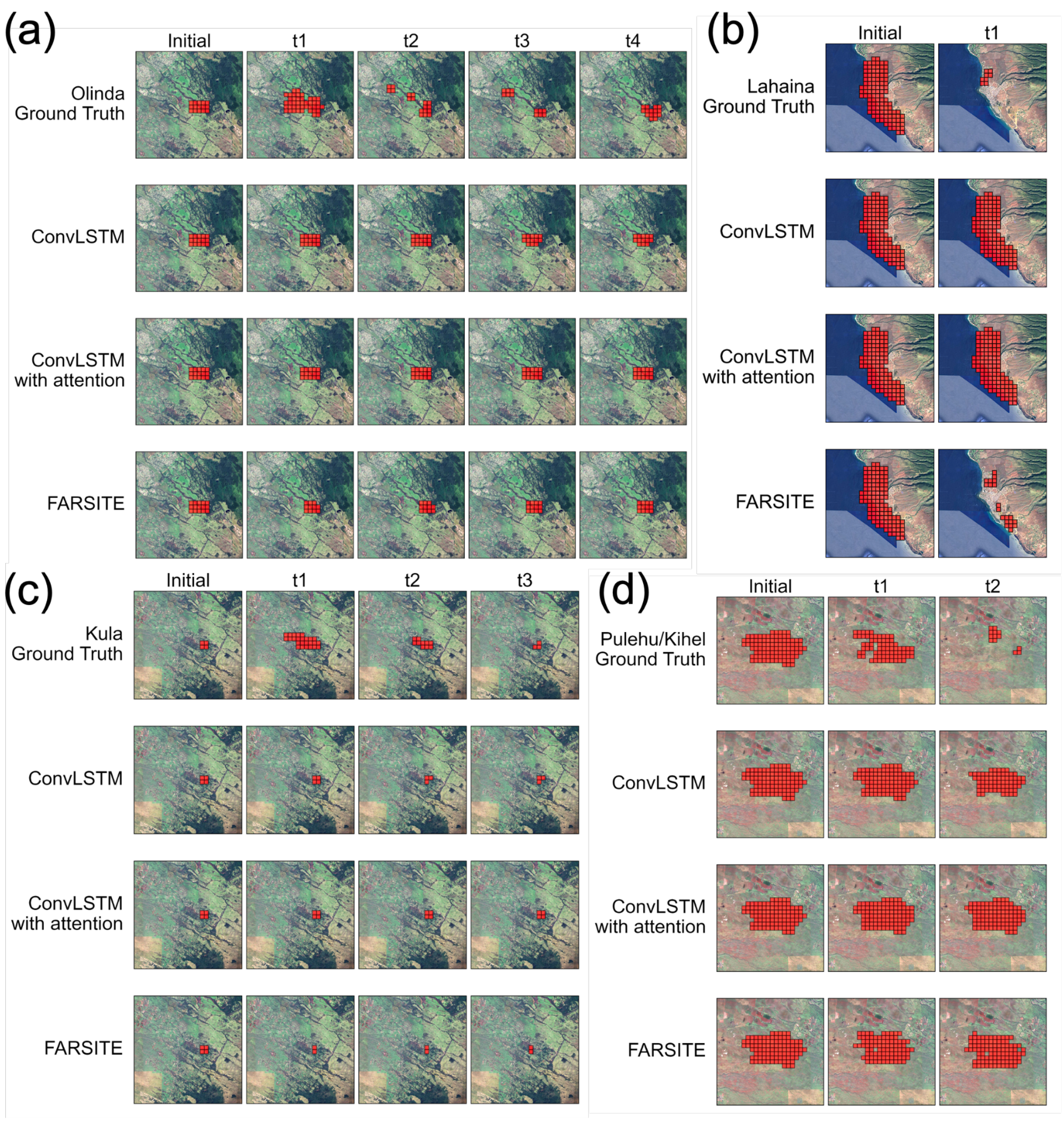

Figure 9. Multi-step prediction results of all models based on the 2023 Maui wildfires.

*4.3 Feature importance for predicting fire spread in the 2023 Maui fires*

The deep learning models can output feature importance to help identify input features that are important for the models to make predictions. Here, we use an explainable AI method, *Integrated Gradients* (Sundararajan et al., 2017), to assess the importance of input features. This method assigns importance scores to input features by calculating the partial derivatives of the model's output with respect to each input feature. Input features with higher gradient values are assigned higher importance scores indicating greater influence of these features on model predictions. We use the Python package *Captum* (Kokhlikyan et al., 2020) to implement this explainable AI method, and this analysis is conducted by focusing on pixels where the model has made correct predictions (i.e., true positives and true negatives).

Figure 10 shows the feature importance averaged across the two deep learning models in different scenarios. The first column shows the feature importance based on the original input data, and the second column shows the feature importance based on the FARSITE-specific input data. The first row shows the feature importance for single-step prediction, and the second row shows the importance for multi-step prediction.

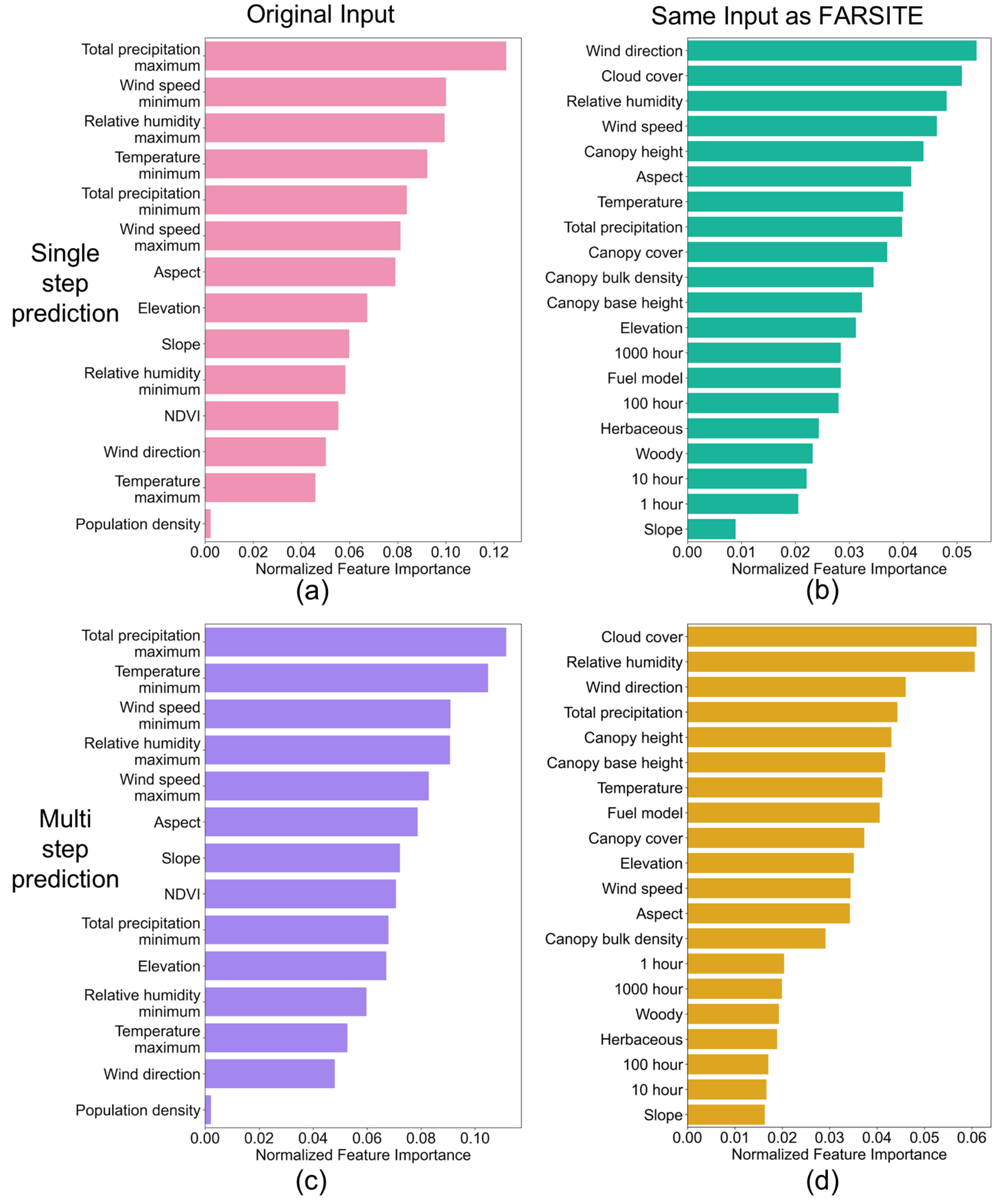

Figure 10. Average feature importance output by the two deep learning models for predicting fire spread in the 2023 Maui fires.

Comparing the two columns in Figure 10, we can see that weather related factors, such as precipitation, wind speed, relative humidity, and temperature, are consistently considered as highly important for predicting fire spread in the 2023 Maui fires, regardless of whether we use the original input data or the input data required by FARSITE. The FARSITE-specific data contain information about canopy height and canopy cover which are also considered among the top important factors. Topographic factors, such as elevation and aspect, are considered moderately important, with feature importance ranked in the middle. By comparing the two rows in Figure 10, we can see that most input features maintain their relative importance in ranking with only minor changes when the models move from single-step prediction to multi-step prediction. Meanwhile, it seems that information about fuel models becomes more important when the models make multi-step predictions than in a single-step prediction. When the original input data is used, the ranking of the input feature *NDVI* moves up from the 11th in single-step prediction to the 8th in multi-step prediction. When the FARSITE-specific input data is used, the ranking of the input feature *fuel model* moves up from the 14th to the 8th. This result suggests that fuel characteristic is more important for the models to make fire spread predictions in longer time steps than in a single step. Population density is shown as the least important feature in both single-step and multi-step predictions. While population density has been used in other deep learning-based wildfire research (Huot et al., 2022; Marjani et al., 2023; Fitzgerald et al., 2023), its usefulness for predicting fire spread at 12-hour intervals can be limited. Other relevant information, such as land use types, could also be used as input features.

## 5 Discussion

### *5.1 Advantages and limitations of different deep learning models*

This study sets out to understand the advantages and limitations of different deep learning models for the task of fire spread prediction. We choose five deep learning models typically used in the literature, i.e., *LSTM*, *U-Net*, *U-Net with attention*, *ConvLSTM*, and *ConvLSTM with attention*, and train and test them on over ten years of wildfire data in the state of Hawaii. For RQ1: *What are the advantages and limitations of five different deep learning models typically used for fire spread prediction?* We find that the *ConvLSTM* model has the highest F1-score among the five tested models, although its F1-score is not significantly higher than the *ConvLSTM with attention* model. The *ConvLSTM* model has a lower recall and a higher precision than the *ConvLSTM with attention* model. Given that a lower recall indicates that the model has missed locations where the fire actually spreads in near future, one could favor *ConvLSTM with attention* that is less likely to miss true fire locations. Compared with the two *ConvLSTM* models, the LSTM model has similar precision but lower recall, while the two *U-Net* models have slightly higher recall but lower precision. This performance pattern generally holds for model predictions in all time steps (see Table 2) and also individual time steps (see Figure 4). Model performance difference is smaller for the first prediction step but becomes larger as the models predict multiple time steps forward. In terms of computational efficiency, the two *ConvLSTM* models require the longest but still acceptable training time, i.e., 10.5 minutes and 17.6 minutes, respectively. The *LSTM* model uses about 2.6 minutes for training, while the two *U-Net* models use the shortest amount of time, i.e., about 1 minute each. The prediction times of all models are within 1 second. The training and test time, as well as the time difference among models, could become larger if a study focuses on a larger study area, e.g., the entire United States. If computational efficiency is not a problem, then training the *ConvLSTM* model or the *ConvLSTM with attention* model may be the best option for predicting fire spread. If computational efficiency is a concern, then one could choose to train the *LSTM* model if favoring precision, or could choose to train one of the *U-Net* models if favoring recall. When using the *LSTM* or *U-Net* models, one could also choose to focus on the predictions in the first one or two steps only to mitigate the effects of decreasing prediction accuracy over longer time steps.

*5.2 Advantages and limitations of deep learning models and FARSITE*

To answer RQ2: *What are the advantages and limitations of the deep learning models compared with a physics-informed and semi-empirical fire spread model?* We have compared the two best deep learning models with the FARSITE model based on the 2023 Maui fire data. We have conducted comparison experiments based on both the original input data used for the other deep learning models as well as the input data specifically prepared based on the requirements of FARSITE. We have also conducted comparison experiments in both single-step prediction and multi-step prediction settings. Overall, we find that FARSITE has higher precision, lower recall, and higher F1-score than *ConvLSTM* and *ConvLSTM with attention* in both single-step and multi-step predictions. Meanwhile, removing the special case of the Lahaina fire (in which FARSITE cannot simulate fire spread on urban lands) leads to similar F1-scores across the three models. While demonstrating better performance in general, FARSITE requires detailed vegetation data, such as fuel model, canopy base height, and canopy height, as well as fuel moisture data to predict fire spread. These data have been made available in the U.S. through LANDFIRE and other data sharing initiatives. In other countries, especially developing countries, however, obtaining such detailed vegetation and fuel moisture data can be difficult. Deep learning models offer more flexibility for the input data, and our experiments show that deep learning models using NDVI as part of the input can achieve a similar performance compared with using detailed vegetation and fuel moisture data required by FARSITE. NDVI data is widely available throughout the world via satellite data products, and this flexibility allows deep learning models to be applied to countries and regions where such detailed data are not available. In terms of computational efficiency, the two deep learning models take between 13.9 and 24.2 minutes to train based on the input data required by FARSITE, but they both take less than 1 second to make predictions. The FARSITE model does not require training time, but it takes more than 1 minute to make predictions. Overall, the computational efficiency of both types of models does not seem to be a concern, and the availability of input data seems to be one major factor for deciding which model to use.

*5.3 Major weather and environmental factors associated with the 2023 Maui wildfires*

By leveraging deep learning models and an explainable AI method (i.e., integrated gradients), we further answer RQ3: *What are the major weather and environmental factors, such as wind,* *vegetation, and topography, associated with the 2023 Maui fires?* Our analysis identifies important factors that both align with and complement those discussed in news and research reports. We find that input variables related to humidity, precipitation, and wind are considered highly important by the deep learning models for predicting the spread of the four Maui fires. This result aligns with existing reports which showed that the drought conditions and strong winds are two major factors driving the spread of Maui fires (Chow, 2023; Partyka & Erdman, 2023; Pequeño, 2023). Our analysis also shows that input variables related to vegetation characteristics, such as canopy height, canopy base height, and canopy cover, are important for predicting fire spread. The importance of these vegetation characteristics could be linked to the flammable shrubs and grasses often from invasive species that fueled the Maui fires (Bushard, 2023; Kluger, 2023). In addition, our analysis shows that topographic factors, such as aspect and elevation, also played fairly important roles in helping the deep learning models predict the locations where the wildfires will spread in the next time steps. The relative importance of factors identified in our study has both similarities and differences in comparison with those identified in studies focusing on other geographic regions of the world. For example, Shadrin et al. (2024) and Khanmohammadi et al. (2022) used machine learning models to predict forest fire spread in northern Russian regions and grass fire spread in Australia, respectively. They both found that wind and vegetation/fuel were the most important factors for predicting wildfire spread; meanwhile, Shadrin et al. (2024) found temperature to be less important for fire spread prediction in northern Russian regions, while Khanmohammadi et al. (2022) did not include slope as an input factor given their focus on grass fires that typically happen on flat or gently undulating terrains. These previous studies, along with our study, collectively contribute to the understanding of factors associated with wildfire spread under different environmental conditions.

*5.4 Limitations and future research*

This study is not without limitations. First, the input weather variables have a coarse spatial resolution (i.e., 9 kilometers), and using variables with higher spatial resolutions will likely help the models better predict wildfire spread. In this research, the study time period is from early 2012 to August 2023, which makes it difficult to use some higher spatial resolution datasets, such as the NAM-Hawaii nest and HRRR-Hawaii nest, that are available for more recent years only (e.g., 2019 and later). We provide a summary of these higher spatial resolution datasets for Hawaii in Supplementary Table S2. Studies that focus on more recent years may leverage these datasets. Another possible way to obtain higher resolution weather data is via downscaling. While this is a complex process that needs to consider atmospheric and environmental factors related to each meteorological variable, properly downscaled data may help further increase the prediction accuracy of wildfire spread. Second, while we have compared five deep learning models and the FARSITE model based on fire data in Hawaii, it could be useful to further compare these models in other geographic regions of the world. To this end, fire datasets in other regions, such as the Portuguese Large Wildfire Spread database (PT-FireSprd) (Benali et al., 2023) and the Canadian Fire Spread Dataset (Barber et al., 2024), could be leveraged to test these models and understand their performance under different weather, fuel, and environmental conditions. Third, while this study has used a variety of input variables, other data, when becoming available, could be added to improve fire spread prediction. For example, some of the fires, whose data are used to train and test models in this study, were likely being fought by firefighters; yet, firefighting data are usually not available. If made available, deep learning models could use such data to distinguish between fire spread changes due to firefighting activities and those due to natural factors (e.g., precipitation or the burning out of fuels), thereby improving model prediction accuracy.

## 6 Conclusions

In this study, we assess five different deep learning models for fire spread prediction based on over ten years of wildfire data in the state of Hawaii. We further compare the best deep learning models with a physics-informed and semi-empirical model, FARSITE, based on the 2023 Maui wildfire data. We find that the two *ConvLSTM* models perform the best among the five tested models, and the FARSITE model has higher precision, lower recall, and higher F1-score than the two *ConvLSTM* models based on the FARSITE-specific input data. The deep learning models show an advantage in their flexibility in input data, i.e., they achieve similar performance using widely available NDVI data compared with using detailed vegetation and fuel moisture data as required by FARSITE. Such flexibility makes deep learning models applicable to geographic regions where detailed vegetation data are not available. Our analysis using deep learning models and explainable AI also identifies important factors associated with the 2023 Maui fires, such as relative humidity, precipitation, and wind. While this study is not without limitations, we hope that it makes a modest contribution by increasing our understanding of the advantages and limitations of different fire spread models and informing future wildfire research and management practices.

**Supplementary Materials**

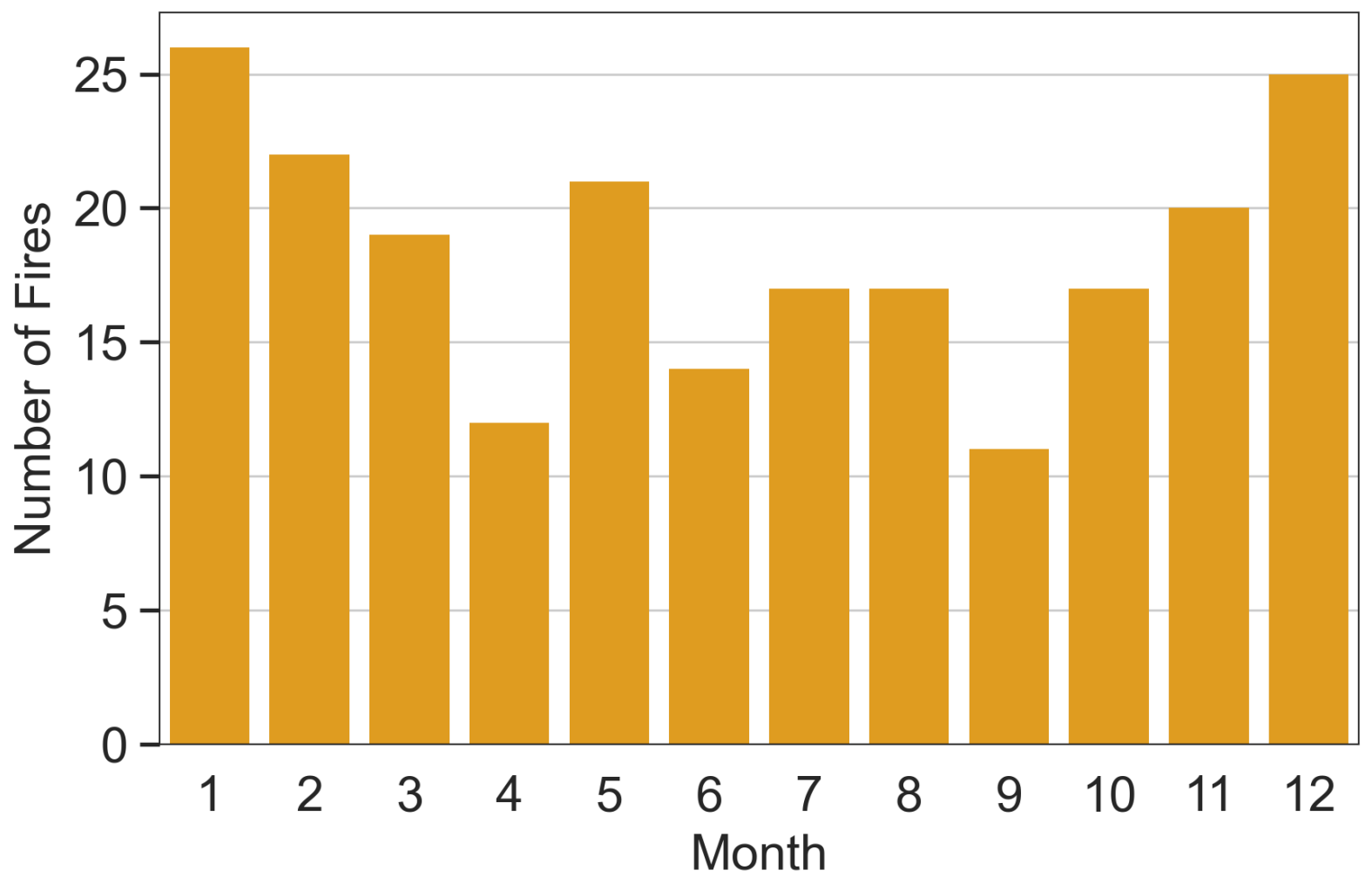

Figure S1. The monthly distribution of wildfires in the dataset.

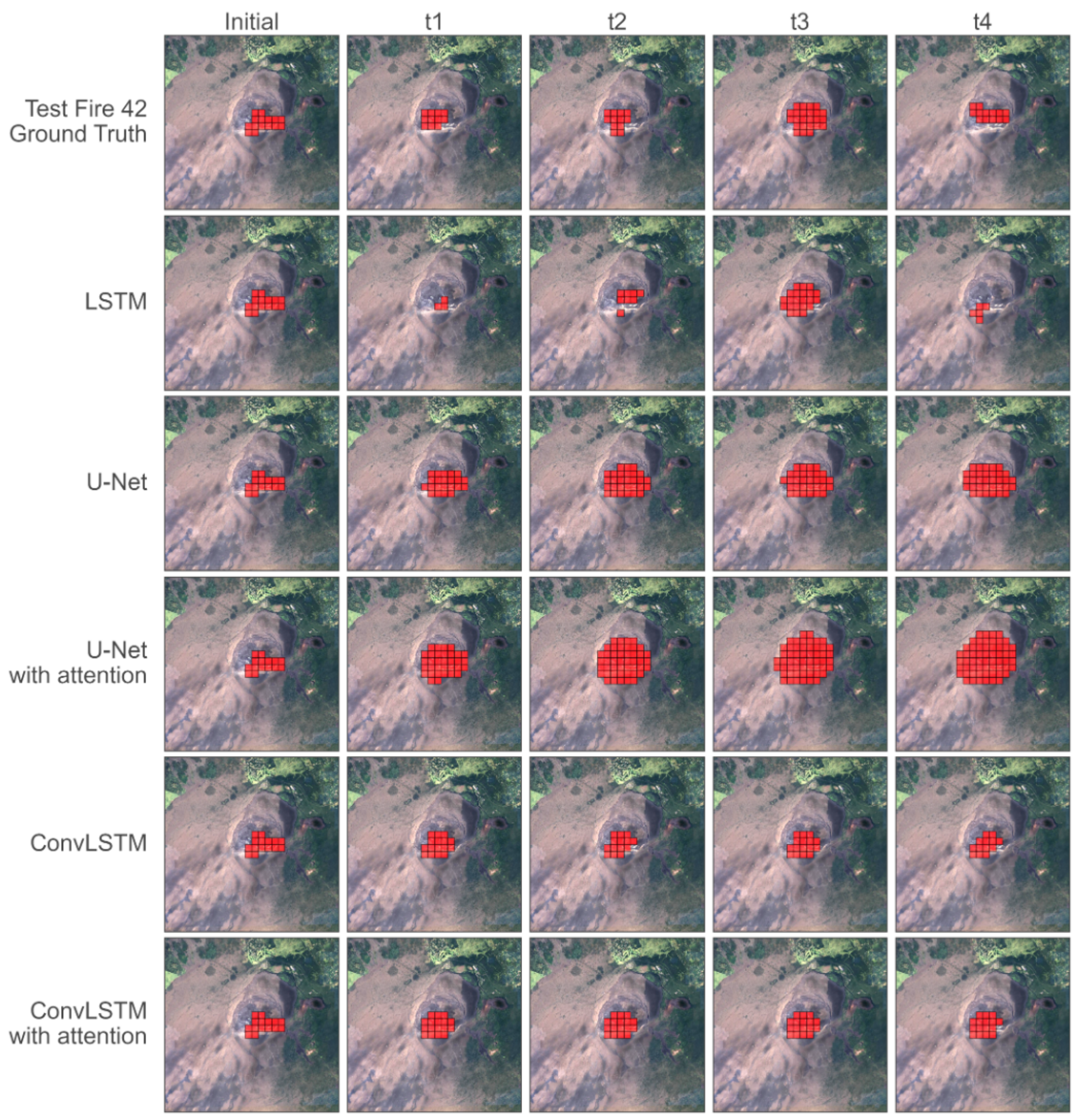

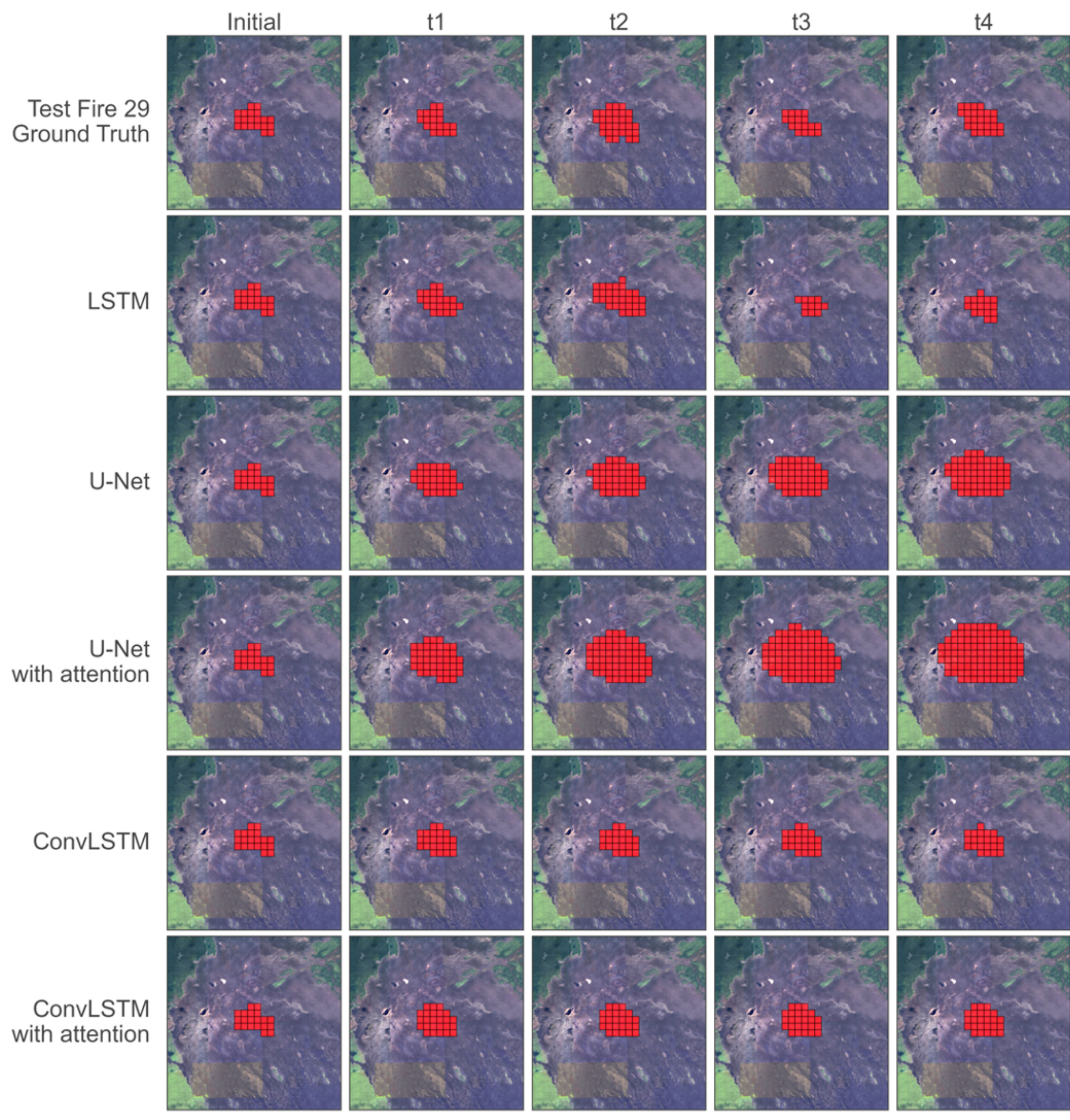

Figure S2. Examples showing the initial fire locations, ground truth fire locations in the next four time steps (12, 24, 36, and 48 hours later), and predicted fire locations from the models.

Table S1. Training and test time of the models based on the same input as FARSITE (unit: second).

| **Model** | **Training time** | **Test time** |
|---|---|---|
| ***ConvLSTM*** | 833.5 | 0.032 |
| ***ConvLSTM with attention*** | 1452.9 | 0.094 |
| ***FARSITE*** | - | 65.200 |

Table S2. Summary of some meteorological datasets with higher spatial resolutions for Hawaii.

| **Data** | **Temporal coverage*** | **Spatial resolution** | **Data availability** |
|---|---|---|---|
| NAM-Hawaii nest | Sep. 2021 - present | 2.5 km | https://noaa-nam-pds.s3.amazonaws.com/index.html (NOAA open data dissemination on AWS) |
| HRRR-Hawaii nest | June 2019 - present | 3 km | HRRR-Hawaii is still at the experimental stage. Currently, HRRR is provided for CONUS and Alaska. |
| NBM-Hawaii | May 2020 - present | 2.5 km | https://noaa-nbm-grib2-pds.s3.amazonaws.com/index.html (NOAA open data dissemination on AWS). Available since Nov. 2016, but publicly released data start from May 2020. |
| HREF-Hawaii | Past 2 days | 3 km | https://nomads.ncep.noaa.gov/pub/data/nccf/com/href/v3.1/ (HREF data repository on NOAA) |
| REFS-Hawaii | Not available | 2.5 km | It is still experimental under intense fine-tuning development as of 2025. |

**Some datasets have different start dates for internal availability and public release. For example, the NBM-Hawaii dataset has been available since November 2016, but the publicly accessible data only start from May 2020. The temporal coverages in this table are based on publicly accessible data.*